\documentstyle[12pt]{article}
\newcommand{\bd}{\begin{document}}
\newcommand{\ed}{\end{document}}
\newcommand{\bc}{\begin{center}}
\newcommand{\ec}{\end{center}}
\newcommand{\be}{\begin{eqnarray}}
\newcommand{\ee}{\end{eqnarray}}
\newcommand{\ba}{\begin{array}}
\newcommand{\ea}{\end{array}}
\newcommand{\eqn}{\global\def\theequation}
\newcommand{\sw}{sin^2 \theta_W}
\newcommand{\fbd}{f_B}
\renewcommand{\thefootnote}{\alph{footnote}}
\newcommand{\se}{\section}
\newcommand{\sse}{\subsection}
\newcommand{\bi}{\bibitem}
\def\figcap{\section*{Figure Captions\markboth
     {FIGURECAPTIONS}{FIGURECAPTIONS}}\list
     {Figure \arabic{enumi}:\hfill}{\settowidth\labelwidth{Figure 999:}
     \leftmargin\labelwidth
     \advance\leftmargin\labelsep\usecounter{enumi}}}
\let\endfigcap\endlist \relax
\def\reflist{\section*{References\markboth
     {REFLIST}{REFLIST}}\list
     {[\arabic{enumi}]\hfill}{\settowidth\labelwidth{[999]}
     \leftmargin\labelwidth
     \advance\leftmargin\labelsep\usecounter{enumi}}}
\let\endreflist\endlist \relax

\begin{document}
\tolerance=10000
\baselineskip=7mm
\begin{titlepage}

 \vskip 0.5in
 \null
\begin{center}
 \vspace{.15in}
{\LARGE {\bf Mesonic Tensor Form Factors with Light Front Quark Model}
}\\
\vspace{1.0cm}
  \par
 \vskip 2.1em
 {\large
  \begin{tabular}[t]{c}
{\bf C.~Q.~Geng$^a$, C. W. Hwang$^{a}$,
 C.~C.~Lih$^{b}$ and W. M. Zhang$^{b}$}
\\
\\
{\sl ${}^a$Department of Physics, National Tsing Hua University}
\\  {\sl  $\ $ Hsinchu, Taiwan, Republic of China }
\\
\\
and
\\
\\
       {\sl ${}^b$Department of Physics, National Cheng Kung University}
\\   {\sl  $\ $Tainan, Taiwan,  Republic of China }\\
   \end{tabular}}
 \par \vskip 5.3em
 {\Large\bf Abstract}
\end{center}

We study the tensor form factors for $P \to P$ and $P \to V$
transitions in the light-front quark model with $P$ and $V$ being
pseudoscalar and vector mesons, respectively.
We explore the behaviors of these form factors in the entire physical
range of $0\leq p^2\leq (M_i - M_f)^2$.
At the maximum recoil of $p^2=0$, we compare our results of the
form factors in $B\to \pi,K,\rho,K^*$ with various
other calculations in the literature.


\end{titlepage}

\se{Introduction}

\ \ \

The studies of the heavy to light exclusive decays in the standard model
are very interesting topics since they can provide the signal
of CP violation and a window into new physics.
Meanwhile, the calculations of these decay widths with the model-independent
way are difficult tasks because they are related to the
non-perturbative QCD effects. These effects correspond to the
binding of quarks in hadrons and appear in matrix elements
of the weak Hamiltonian operators between initial and final hadronic states.
However, they could be described by form factors
from the parametrization of those hadronic matrix elements in
Lorentz invariant. Phenomenologically, these form factors can be
evaluated in many different approaches, such as the lattice QCD, the QCD
sum rule,
the non-relativistic quark model and the light-front quark model (LFQM).

It is well known that when the momentum of the final state meson
increases $(>1 GeV)$, we have to
consider the relativistic effect seriously, especially, at the
maximum recoil of $p^2=0$ where the final meson could be highly
relativistic, and
there is no reason to expect that the non-relativistic quark model is
still applicable. While the lattice QCD is still not practically useful
for a calculation in hadron physics,
the LFQM \cite{ter,wmz,jau} is the only relativistic quark model in which
quark spins and the center-of-mass motion can be carried out in a
consistent and fully relativistic treatment. The light-front wave
function is manifestly boost invariant as it is expressed in terms of the
longitudinal momentum fraction
and relative transverse momentum variables.
The parameter in the hadronic wave function is
determined from other information and the meson state of the definite
spins could be constructed by the Melosh transformation \cite{melosh}.

The LFQM has been widely applied to
study the form factors of weak decays
\cite{Don94,Don95,cheng1,cheng2,geng1,geng2,jau2}.
In the most pervious works with the LFQM \cite{Don95,jau2}, the
$P \to P$ and $P \to V$ ($P$=pseudoscalar meson, $V$=vector meson)
tensor form factors were calculated only
for $p^2 \leq 0$. However, physical decays
occur in the timelike region $0\leq p^2\leq (M_i - M_f)^2$ with $M_{i,j}$
being the initial and final meson masses, respectively. Hence,
to extrapolate the form factors to cover the entire range of the momentum
transfer,  some assumptions are needed.
For example, in Ref. \cite{Jaus96}, a light front ansatz for the $p^2$
dependence was made to
extrapolate the form factors in the space-like to the time-like regions
and in Ref. \cite{Mel}, based on the dispersion formulation, form factors
at $p^2>0$ were obtained
by performing an analytic continuation from the space-like
$p^2$ region. In Refs. \cite{cheng1,Sima,Simb}, for the first time the
form factors in the $P\to P$ transition were calculated for the entire
range of $p^2$ in the LFQM. In this work, we calculate the tensor form
factors for both $P \to P$ and $P \to V$ transitions in the entire range
of the momentum transfer
$0\leq p^2\leq (M_i - M_f)^2$.
These tensor form
factors play an important role in estimating the B decay rates 
 such as that of
$B\to K^{(*)}l^+l^-$ \cite{geng-B}.

This paper is organized as follows. In Sec.~2,
we study the tensor form factors of $P \to P$ and $P \to V$
transitions within the framework of the LFQM. In Sec.~3, we present our
numerical results. We give the conclusions in Sec.~4, .

\se{Tensor Form Factors}

\ \ \

\sse{Framework}

We start with the weak tensor operator, given by:
\be
O^{\mu \nu}=\bar q i \sigma^{\mu \nu} p_{\nu}(1+\gamma_5) Q
\label{top}
\ee
where $\bar q$ is the light anti-quark. Our main task is to evaluate
the tensor form factors for $P \to P$ and $P \to V$ transitions. They are
defined by the following hadronic matrix elements:
\be
\langle P_2(P_{f})|\bar{q}i \sigma^{\mu \nu} p_{\nu}Q|P_1(P_{i})\rangle &=&
\frac{F_T(p^2)}{M_{P_1}+M_{P_2}}[(P_i+P_f)^{\mu}p^2
-(M_{P_{1}}^{2}-M_{P_2}^{2})p^{\mu}] \, , \nonumber\\
\langle V(P_f,\epsilon)|\bar{q}i \sigma^{\mu \nu} p_{\nu}Q|P(P_i)\rangle &=&
-i\varepsilon^{
\mu\nu\alpha\beta}\epsilon^*_{\nu} P_{f \alpha} P_{i\beta} F_{1}(p^2) ,
 \nonumber\\
\langle V(P_f,\epsilon)|\bar{q}i \sigma^{\mu \nu} p_{\nu}\gamma_5
Q|P(P_i)\rangle &=&
\left[(M_{P}^{2}-M_{V}^{2})\epsilon^{*\mu}
-(\epsilon^* \cdot p) (P_f+P_i)^{\mu}\right]F_{2} \nonumber\\
&&+\epsilon^* \cdot p\left[p^{\mu}
-\frac{p^2}{M_{P}^{2}-M_{V}^{2}}(P_f+P_i)^{\mu}\right]F_{3}(p^2) ,
\label{tff}
\ee
where $P_{i(f)}$ is the momentum of
 the initial (final) state meson, $\epsilon$ is the meson polarization
vector and $p=P_i-P_f$.

In calculations of the hadronic matrix elements,
one usually lets $p^+=0$ which leads to a spacelike momentum transfer.
However, since the momentum transfer should be always timelike in a
real decay process, in this work, the tensor form factors in Eq.~(\ref{tff})
will be calculated in the frame of $p_\perp=0$, namely the physically
accessible kinematic region $0\leq p^{2}\leq p^2_{\max }$. In the LQFM,
the meson bound state, which consists of a quark $q_1$ and an anti-quark
$\bar{q_2}$ with the total momentum $P$ and spin $S$, can be written as
\begin{eqnarray}
|M(P,S,S_z)\rangle &=& \int [dk_{1}][dk_{2}]
2(2\pi)^{3}\delta ^{3}(P-k_{1}-k_{2}) \nonumber \\
&\times&\sum_{\lambda _{1}\lambda_{2}}
\Phi^{S S_{z}}(k_1,k_2,\lambda_1,\lambda_2)
|q_1(k_{1},\lambda_{1}) \bar{q}_2( k_{2},\lambda_2)
\rangle\,,
\label{mwf}
\end{eqnarray}
where
\be
&&[dk]={\frac{dk^+d^2k_{\bot}}{2(2\pi)^3}}\,,  \nonumber \\
&&|q_1(k_{1},\lambda _{1}) \bar{q}_2( k_{2},\lambda_{2})
\rangle = b_{q_1}^\dagger(k_{1},\lambda_{1}) d_{\bar{q}_2}^\dagger
( k_{2},\lambda_{2})|0 \rangle, \nonumber \\
&&\{b_{\lambda'}(k'),b^\dagger_{\lambda}(k)\}
=\{d_{\lambda'}(k'),d^\dagger_{\lambda}(k)\}
=2(2\pi)^3~\delta^3(k'-k)~\delta_{\lambda'\lambda}.
\ee
and $k_{1(2)}$ is the on-mass shell light front momentum of the internal
quark $q_{1}(\bar{q_2})$. The light-front relative momentum variables
$(x,k_{\bot})$ are defined by
\begin{eqnarray}
        && k^+_1=x_1 P^+, \quad k^+_2=x_2 P^+, \quad x_1+x_2=1, \nonumber \\
        && k_{1\bot}=x_1 P_\bot+k_\bot, \quad k_{2\bot}=x_2
                P_\bot-k_\bot,
\end{eqnarray}

In the momentum space, the wave function $\Phi^{SS_z}$ can be written as:
\begin{equation}
        \Phi^{SS_z}(k_1,k_2,\lambda_1,\lambda_2)
                = R^{SS_z}_{\lambda_1\lambda_2}(x_i,k_\bot)~ \phi(x_i, k_\bot),
                \label{distribution}
\end{equation}
where $\phi(x,k_{\bot})$ describes the momentum
distribution amplitude of the constituents in the bound state,
$R^{SS_z}_{\lambda_1\lambda_2}$ constructs a spin state $(S,S_z)$ out
of light-front helicity eigenstates $(\lambda_1\lambda_2)$ which could be
expressed as:
\be
R^{SS_z}_{\lambda_1 \lambda_2}(x_i,k_\bot)
                =\sum_{s_1,s_2} \langle \lambda_1|
                {\cal R}_M^\dagger(x_1,k_\bot, m_1)|s_1\rangle
                \langle \lambda_2|{\cal R}_M^\dagger(x_2,-k_\bot, m_2)
                |s_2\rangle
                \langle {1\over2}s_1
                {1\over2}s_2|SS_z\rangle,
\label{n6}
\ee
where $|s_i\rangle$ are the Pauli spinors,
and ${\cal R}_M$ is the Melosh transformation operator:
\be
        {\cal R}_M (x,k_\bot,m_i) =
                {m_i+x_iM_0+i\vec \sigma\cdot\vec k_\bot \times \vec n
                \over \sqrt{(m_i+x_i M_0)^2 + k_\bot^2}},
\ee
with
\be
        M_0^2={ m_1^2+k_\bot^2\over x_1}+{ m_2^2+k_\bot^2\over x_2}\, ,
\label{M0}
\ee
and $\vec{n}=(0,~0,~1)$.
Actually, if we set $x_1=1-x$ and $x_2=x$, Eq.~(\ref{n6}) can be given by
a covariant form \cite{jau}.
\be
 R^{SS_z}_{\lambda_1\lambda_2}(x,k_\bot)
                ={\sqrt{k_1^+k_2^+}\over \sqrt{2} ~{\widetilde M_0}}
        ~\bar u(k_1,\lambda_1)\Gamma v(k_2,\lambda_2),,
\label{rs}
\ee
with
\be
&&{\widetilde M_0} \equiv \sqrt{M_0^2-(m_1-m_2)^2}, \nonumber\\
&& \sum_\lambda u(k,\lambda) \overline{u}(k,\lambda) = {\frac{m + \not{\!
}k }{k^+}} \, , \ \ \sum_\lambda v(k,\lambda) \overline{v}(k,\lambda) = {%
\frac{m - \not{\! }k }{k^+}} \, ,
\ee
where $\Gamma $ stands for
\be
&&\Gamma=\gamma_5 \qquad ({\rm pseudoscalar}, S=0),
\nonumber\\
&&\Gamma=-\not{\! \hat{\varepsilon}}(S_z)+
          {\hat{\varepsilon}\cdot(p_1-p_2)
                \over M_0+m_1+m_2} \qquad ({\rm vector}, S=1),
\ee
and
\begin{eqnarray}
        &&\hat{\varepsilon}^\mu(\pm 1) =
                \left[{2\over P^+} \vec \varepsilon_\bot (\pm 1) \cdot
                \vec P_\bot,\,0,\,\vec \varepsilon_\bot (\pm 1)\right],
                \quad \vec \varepsilon_\bot
                (\pm 1)=\mp(1,\pm i)/\sqrt{2}, \nonumber\\
        &&\hat{\varepsilon}^\mu(0)={1\over M_0}\left({-M_0^2+P_\bot^2\over
                P^+},P^+,P_\bot\right).   \label{polcom}
\end{eqnarray}
The normalization condition of the meson state is given by
\be
&&\langle M(P',S',S'_z)|M(P,S,S_z)\rangle = 2(2\pi)^{3} P^{+}
\delta^{3}(\tilde P'- \tilde P)\delta_{S'S}\delta_{S'_zS_z} \, ,
\ee
which leads to
\be
\int {dx\, d^2k_\bot\over 2(2\pi)^3} |\phi(x,k_\bot)|^2 = 1\, .
\ee

In principle, the momentum distribution amplitude $\phi(x,k_\bot)$
can be obtained by solving the light-front QCD bound state equation
\cite{wmz2,wmz3}. However, before such first-principle solutions are
available, we would have to be contented with
phenomenological amplitudes. One example that has been often used in the
literature is the Gaussian type wave function:
\be
\phi(x,k_{\bot})=N\sqrt{\frac{dk_{z}}{dx}}
\exp \left( -\frac{\vec{k}^{2}} {2\omega_{M}^{2}}\right) \,,
\label{7}
\ee
where $N = 4 \left({\frac{\pi}{\omega_{M}^{2}}}\right)^{\frac{3}{4}}$,
$\vec k = (k_{\bot}, k_z)$, $k_z$ is defined through
\be
1-x = {e_1-k_z\over e_1 + e_2} \, \ \
x = {e_2+k_z \over e_1 + e_2}\, , \ \  e_i = \sqrt{m_i^2 + \vec k^2} \,
\ee
by
\be
& &
\ \ k_{z} =\left( x-\frac{1}{2}\right) M_{0}+\frac{m_{1}^{2}-m_{2}^{2}}{%
2M_{0}} \, .
\ee
and
\be
{{dk_z\over dx}} = \,{e_1 e_2\over x(1-x)M_0}\,
\ee
is the Jacobian of the transformation from $(x,k_{\bot})$ to $\vec k$.
In particular, with appropriate parameters, this wave function in Eq. (16)
describes satisfactorily the pion elastic form factor up to
$p^2\sim 10~{\rm GeV}^2$ \cite{Chung}.

\sse{Decay constants}

In order to use the light front bound states in Eq.~(\ref{mwf}) to calculate the
matrix elements in
Eq.~(\ref{tff}), we need to fix the parameter $\omega$ in the wave
function of Eq.~(\ref{7}). This parameter can be determined by the
corresponding pseudoscalar and vector mesonic decay constants, defined by
$\langle 0|A^\mu|P\rangle=\,if_{_P}P^\mu$ and $\langle 0|V^\mu|V\rangle=
\,f_VM_V\epsilon^\mu$, respectively.

 From  Eq.~(\ref{mwf}), one has
\be
\langle 0|\bar{q}_2\gamma^+\gamma_5 q_1|P\rangle &=& \int [d^3p_1][d^3p_2]
2(2\pi)^3\delta^3(P-p_1-p_2)\phi_P(x,k_\perp)R^{00}_{\lambda_1\lambda_2}
(x,k_\perp)   \nonumber \\
&& \times\,\langle 0|\bar{q}_2\gamma^+\gamma_5 q_1|q_1(p_{1},\lambda_{1})
\bar{q}_2( p_{2},\lambda_2)\rangle,
\ee
It is straightforward to show that
\be
f_P=\,4{\sqrt{3}\over\sqrt{2}}\int {dx\,d^2k_\perp\over 2(2\pi)^3}\,\phi_P(x,
k_\perp)\,{{\cal A}\over\sqrt{{\cal A}^2+k_\perp^2}}, \label{fp}
\ee
where
\be
{\cal A}=m_1x+m_2(1-x).
\ee
Note that the factor $\sqrt{3}$ in (\ref{fp}) arises from the color factor
implicitly in the meson wave function.

   Similarly,  the vector-meson decay constant is found to be
\be
f_V &=& 4{\sqrt{3}\over \sqrt{2}}\int {dx\,d^2k_\perp\over 2(2\pi)^3}\,{\phi
_V(x,k_\perp)\over\sqrt{{\cal A}^2+k^2_\perp}}\,{1\over M_0}\Bigg\{ x(1-x)
M_0^2+m_1m_2+k^2_\perp   \nonumber \\
&& +\,{{\cal B}\over 2 W_V}\left[ {m_1^2+k_\perp^2\over 1-x}-{m_2^2+k_\perp^2
\over x}-(1-2x)M_0^2\right]\Bigg\},   \label{fv}
\ee
where
\be
{\cal B}=xm_1-(1-x)m_2,~~~~W_V=M_0+m_1+m_2.
\ee
If the decay constant is experimentally known, one can determine the
parameter $\omega$ in the light-front wave function.

\sse{Tensor form factor for $P \to P$ transition}

The hadronic matrix element of the tensor operator for the $P \to P$
transition could be newly parametrized in terms of the initial meson
momentum $p+q$ and final meson momentum $q$, that is,
\be
\langle P_{2}(q)|\bar{q}i\sigma^{\mu\nu}p_{\nu}Q|P_{1}(p+q)\rangle =
\frac{F_T}{M_{P_{1}}+M_{P_{2}}}[(p+2q)^{\mu}p^2
-(M_{P_{1}}^{2}-M_{P_{2}}^{2})p^{\mu}] ,
\label{ft}
\ee
where $p$ is the momentum transfer.
For $P_1=(Q_1\bar q)$ and $P_2=(Q_2\bar q)$, the relevant quark momentum
variables are
\be
&&p_{Q_1}^{+}=(1-x)(p+q)^+,
~p^{+}_{\bar q}=x(p+q)^+,~p_{Q_{1\perp}}=(1-x)q_{\perp}+k_\perp,
~p_{\bar q\perp}=xq_{\perp}-{k}_\perp, \nonumber \\
&&p_{Q_2}^{+}=(1-x')q^+,
~~p'^+_{\bar q}=x'q^+,~~p_{Q_{2\perp}}=(1-x')q_\perp+k'_\perp,
~~p'_{\bar q\perp}=x'q_\perp-k'_\perp,  \label{transmom}
\ee
where $x~(x')$ is the momentum fraction of the anti-quark
in the initial (final) state. At the quark loop, this anti-quark is the
spectator, thus it requires that
\be
p'^{+}_{\bar q}=p_{\bar q}^+,~~~~~p'_{\bar q\perp}=p_{\bar q\perp}.
\label{momeq}
\ee
To calculate the matrix element in Eq. (\ref{ft}), we take
a Lorentz frame where the transverse momentum $p_{\bot}$ = $0$
such that $p^{2}=p^{+}p^{-} \geq 0$ covers the
entire physical region of the momentum
transfer \cite{Dubin}.

Consider the ``good" component of $\mu=+$, we have
\be
\langle P_2|\bar{q}i\sigma^{+\nu}p_{\nu}Q|P_1\rangle &=&\sqrt{x \over x'}
\int {dx \,d^2k_\bot \over2(2\pi)^3}\,
\phi^*_{P_2}(x',k_\bot) \phi_{P_1}(x,k_\bot)
{1 \over 2 \widetilde M_{0_{P_2}} \widetilde M_{0_{P_1}}
\sqrt{(1-x')(1-x)} } \nonumber\\
&& \times {\rm Tr}\left[\gamma_5(\not{\! p}_2+m_{Q_2})\sigma^{+\nu}p_{\nu}
(\not{\! p}_1+m_{Q_1}) \gamma_5 (\not{\! p}_{\bar q}-m_{\bar q})\right].
\label{mel3}
\ee
The trace in the above expression can be easily carried out. By using
Eqs.~(\ref{ft}) and (\ref{mel3}),
the form factor $F_T$ is found to be
\be
F_{T}(p^2)=\frac{M_{P_1}+M_{P_{2}}}{(1+2r)p^{2}-(M_{P_1}^{2}-M_{P_{2}}^2)}
\int dx' d^{2}k_{\bot}\frac{\phi^*_{P_2}(x',k_{\bot})
\phi_{P_1}(x,k_{\bot})}{{\widetilde M_{0_{P_2}} \widetilde M_{0_{P_1}}
\sqrt{(1-x')(1-x)}}} A
\ee
where $x=x' r/(1+r)$ and
\be
A&=&\frac{1}{x x'(1-x)(1-x')}
\bigg\{(x m_{Q_1}+(1-x)m_{\bar q})(x' m_{Q_2}+(1-x')m_{\bar q})  \nonumber\\
&&\left[x(1-x')m_{Q_1}-x'(1-x)m_{\bar q}\right]
+k^{2}_\perp \biggl[x'(1-x')(2x-1)m_{Q_2}  \nonumber\\
&&+(x-x')(x+x'-2xx')m_{\bar q}+x(1-x)(1-2'x)m_{Q_1}\biggr]\bigg\}\,.
\ee
At the maximum recoil of $p^2=0$, we have $x=x'$ and get
\be
F_{T}(0)&=&-\frac{1}{M_{P_1}-M_{P_{2}}}
\int dx d^{2}k_{\bot}\phi^*_{P_2}(x,k_{\bot})
\phi_{P_1}(x,k_{\bot})  \nonumber\\
&&\times \frac{1}{x^{2}(1-x)\widetilde M_{0_{P_2}} \widetilde M_{0_{P_1}}}
\bigg\{(x m_{Q_1}+(1-x)m_{\bar q})(x m_{Q_2}+(1-x)m_{\bar q})  \nonumber\\
&&\left[x(1-x)m_{Q_1}-x(1-x)m_{Q_2}\right]
+k^{2}_\perp x(1-x)(1-2x)(m_{Q_1}-m_{Q_2})\bigg\}\,.
\ee

\sse{Tensor form factors for $P \to V$ transition}

Similarly, the tensor form factors $F_{1,2,3}$ for the $P \to V$ transition
are defined as:
\be
\langle V(q,\epsilon)|\bar{q}i\sigma^{\mu\nu}p_{\nu}Q|P(p+q)\rangle &=&
-i\varepsilon^{
\mu\nu\alpha\beta}\epsilon^*_\nu q_\alpha p_\beta F_{1} ,
 \nonumber\\
\langle V(q,\epsilon)|\bar{q}i\sigma^{\mu\nu}p_{\nu}\gamma_5
Q|P(p+q)\rangle &=&
\left[(m_{P}^{2}-m_{V}^{2})\epsilon^{*\mu}
-(\epsilon^* \cdot p) (p+2q)^{\mu}\right]F_{2} \nonumber\\
&&+\epsilon^* \cdot p\left[p^{\mu}
-\frac{p^2}{m_{P}^{2}-m_{V}^{2}}(p+2q)^{\mu}\right]F_{3} ,
\label{5}
\ee
where $q$ and $p+q$ are the four momenta
of the vector and pseudoscalar mesons,
and $\epsilon^{\mu }$ is the meson polarization
vector, given by
\be
\varepsilon^\mu(\pm 1) =
                \left[{2\,\vec \varepsilon_\bot \cdot \vec q_\bot\over q^+}
,\,0,\,\vec \varepsilon_\bot \right],
                \quad
\varepsilon^\mu(0)={1\over M_V}\left({-M_V^2+q_\bot^2\over
                q^+},q^+,q_\bot\right)\,,
\ee

respectively.
The calculation of the $P \to V$ form factors is more subtle
than that in the $P \to P$ case. Here we use the two-component
form of the light-front quark field \cite{wmz4}.
For the ``good" component of $\mu=+$, the tensor current
in Eq.~(\ref{top}) can be written as:
\be
\bar{q}i\sigma^{+\nu}p_{\nu}(1+\gamma_5) Q=\{q_{+}^{\dagger}\gamma^{0}
(1+\gamma_5) Q_{-}-
q_{-}^{\dagger}\gamma^{0}(1+\gamma_5)Q_{+}\}p^+
\label{op1}
\ee
where the $q_{+}(Q_{+})$ and $q_{-}(Q_{-})$ are the light-front up
and down components of the quark fields, expressed by \cite{wmz4}
\be
q_{+}&=&\left(\ba{c}\chi_{q} \\ 0\ea\right) \,
\ee
and
\be
q_{-}&=&\frac{1}{i\partial^{+}}(i\alpha_{\bot}\cdot
\partial_{\bot}+\beta m_{q})q_{+}
  = \left(\ba{c}0 \\ \frac{1}{\partial^{+}}
(\tilde{\sigma}_{\bot} \cdot\partial_{\bot}+m_{q})
\chi_{q}\ea\right) \, ,
\ee
respectively. Then, the tensor operator in the
Eq. (\ref{op1}) could be written as follow:
\be
&&q_{+}^{\dagger}\gamma^{0}(1+\gamma_{5})Q_{-} =
-i\,\chi_{q}^{\dagger}(1-\sigma_{3})\frac{1}{\partial^{+}}
(\tilde{\sigma}_{\bot} \cdot \stackrel{\rightarrow}
\partial_{\bot}+m_{Q})\chi_{Q} \, , \nonumber\\
&&q_{-}^{\dagger}\gamma^{0}(1+\gamma_{5})Q_{+} =
i\,[\frac{1}{\partial^{+}}\chi_{q}^{\dagger}
(\tilde{\sigma}_{\bot} \cdot  \stackrel{\leftarrow} \partial_{\bot}+m_{q})]
(1+\sigma_{3})\chi_{Q} \, ,
\label{op}
\ee
where $\chi_{q(Q)}$ is a two-component spinor field
and $\sigma$ is the Pauli matrix.

Let $P=(Q_1 \bar q)$ and $V=(Q_2 \bar q)$.
By the use of Eqs.~(\ref{mwf}), (\ref{op1}) and (\ref{op}),
Eq.~(\ref{5}) can be reduced to
\be
&&\langle V(q,\epsilon)|\bar{q}i\sigma^{+\nu}p_\nu Q |P(p+q)\rangle  
\nonumber\\
&&~~~~=-i\int [dk_{1}][dk'_{1}][dk_{2}]2(2\pi)^{3}\delta^{3}(p+q-k_{1}-k_{2})
\,2(2\pi)^{3}\delta^{3}(q-k'_{1}-k_{2})  \nonumber\\
&&~~~~~~\times \sum\Phi_{P}^{\lambda _{1}\lambda _{2}}(k_{1},k_{2},\lambda_{1},
\lambda_{2})
\sum\Phi_{V}^{* \lambda'_{1}\lambda _{2}}(k'_{1},k_{2},\lambda'_{1},
\lambda_{2})
\nonumber \\
&&~~~~~~\times\chi_{\lambda'_{1}}^{+}\bigg\{\frac{1}{k_{1}^{+}}
(\tilde{\sigma}_{\bot} \cdot k_{1_{\bot}}+im_{Q})+\frac{1}{k_{1}^{+'}}
(\tilde{\sigma}_{\bot} \cdot k_{1_{\bot}}'+im_{q})
\bigg\}\chi_{-\lambda'_{2}} \, , \nonumber \\
&&\langle V(q)|\bar{q}i\sigma^{+\nu}p_\nu \gamma_5 Q |P(p+q)\rangle 
 \nonumber\\
&&~~~~=-i\int [dk_{1}][dk'_{1}][dk_{2}]2(2\pi)^{3}\delta^{3}(p+q-k_{1}-k_{2})
\,2(2\pi)^{3}\delta^{3}(q-k'_{1}-k_{2})  \nonumber\\
&&~~~~~~~\times \sum\Phi_{P}^{\lambda _{1}\lambda _{2}}(k_{1},k_{2},
\lambda_{1},\lambda_{2})
\sum\Phi_{V}^{* \lambda'_{1}\lambda _{2}}(k'_{1},k_{2},\lambda'_{1},
\lambda_{2})
\nonumber \\
&&~~~~~~~\times\chi_{\lambda'_{1}}^{+}\bigg\{\sigma_{3}\frac{-1}{k_{1}^{+}}
(\tilde{\sigma}_{\bot} \cdot k_{1_{\bot}}+im_{Q})+\frac{1}{k_{1}^{+'}}
(\tilde{\sigma}_{\bot} \cdot k_{1_{\bot}}'+im_{q})
\sigma_{3}\bigg\}\chi_{-\lambda'_{2}} \, .
\label{loop}
\ee
The form factor $F_{1}$ is only related to the $1^-$ intermediate state
with the vector daughter meson but $F_2$ and $F_3$ mix with $1^+$ and
$0^+$ states. Therefore, we have to calculate the matrix elements
for $1^+$ and $0^+$ states with different vector meson polarizations.
For $\varepsilon^\mu(\pm 1)$ and
$\varepsilon^\mu(0)$, $R^{SS_z}_{\lambda_1\lambda_2}(x,k_\bot)$ in
$\Phi_{V}^{* \lambda'_{1}\lambda _{2}}(k'_{1},k_{2},\lambda'_{1},\lambda_{2})$
can be written respectively as:
\be
R^{SS_z}_{\lambda_1\lambda_2}(x,k_\bot)&=&
{\sqrt{p_1^+p_2^+}\over \sqrt{2} ~{\widetilde M_{0V}}}\chi^{+}_{\lambda_1}
\bigg\{
\frac{(1-x)\sigma_{\bot} \cdot P_{\bot}+\sigma_{\bot} \cdot k_{\bot}-im_{Q_2}}
{(1-x)P^+}\sigma_{\bot} \cdot \epsilon_{\bot} \nonumber\\
        &&-2\frac
{\epsilon_{\bot}\cdot P_{\bot}}{P^+}+\sigma_{\bot} \cdot \epsilon_{\bot}
\frac{x\sigma_{\bot} \cdot P_{\bot}-\sigma_{\bot}
 \cdot k_{\bot}-im_{\bar q}}{xP^+} \nonumber\\
        &&+2\frac{\epsilon_{\bot}\cdot k_{\bot}}
{W_{V}x(1-x)P^+} [-i\sigma_{\bot} \cdot k_{\bot}-xm_{Q_2}+(1-x)m_2]\bigg\}
\chi_{-\lambda_2}
\label{p1}
\ee
and
\be
R^{SS_z}_{\lambda_1\lambda_2}(x,k_\bot)&=&
{\sqrt{p_1^+p_2^+}\over \sqrt{2} ~{\widetilde {M}_{0V}}}
\frac{1}{M_{0V}P^{+}}\chi^{+}_{\lambda_1}\bigg\{M_{0V}^{2}-P_{\bot}^{2}
 \nonumber\\
        &&+\frac{1}{x(1-x)}[\,x(1-x)(\sigma_{\bot} \cdot P_{\bot})
(\sigma_{\bot} \cdot P_{\bot}) \nonumber\\
        &&+(\sigma_{\bot} \cdot k_{\bot})
(\sigma_{\bot} \cdot k_{\bot})-i(m_{Q_2}-m_{\bar q})
(\sigma_{\bot} \cdot k_{\bot})
+m_{Q_2}m_{\bar q}\,] \nonumber\\
        &&+\frac{1}{2W_{V}x(1-x)}[\,\frac{m_{Q_2}^{2}+k_{\bot}^{2}}{1-x}
-\frac{m_{\bar q}^{2}+k_{\bot}^{2}}{x}-(1-2x)M_{0V}^{2}\,] \nonumber\\
        &&
\times [\,i(\sigma_{\bot} \cdot k_{\bot})+xm_{Q_2}-(1-x)m_{\bar q}\,]
\bigg\}
\chi_{-\lambda_2}
\label{p0}
\ee
where,
\be
W_{V}=M_{0V}+m_{Q_2}+m_{\bar q}\,,~~~M_{0V}^{2}=
{ m_{Q_2}^2+k_\bot^2\over {1-x'}}+{ m_{\bar q}^2+k_\bot^2\over x'}.
\ee

Since the form factor $F_{1}$ is only related to
$\varepsilon^\mu(-1)$, we can find the matrix
element as follow:
\be
\langle V(q,\epsilon)|\bar{q}i\sigma^{+\nu}p_\nu Q |P(p+q)\rangle
=F_{1} \epsilon ^{ij}\epsilon_{i}^{*} q_{j}\,.
\label{f1}
\ee
Using Eqs.~(\ref{loop}) and (\ref{f1}), we obtain
\be
F_{1}=\int dx' d^{2}k_{\bot}N_{V}\phi^{*}_{V}(x',k_{\bot})
N_{P}\phi_{P}(x,k_{\bot})(A_{1}+A_2)\, ,
\label{f1r}
\ee
where
\be
N\phi(x,k_{\bot})&=&4\left(\frac{\pi}{\omega^{2}_{M}}\right)^{\frac{3}{4}}
\frac{1}{\sqrt{2}\sqrt{M_{0}^{2}-(m_Q-m_{\bar{q}})^2}}
\sqrt{\frac{e_Q e_{\bar{q}}}{M_0}}\exp\left(-\frac{\vec{k}^{2}}
{2\omega_{M}^{2}}\right)\, ,
\ee
\be
A_1&=&\frac{2}{xx'^{2}(1-x')^{2}(1-x)^{2}}
\bigg\{(1-2x'+xx')(x'+x-2x'x)k^{2}_{\bot} \nonumber\\
&&+[x'm_{Q_2}+(1-x')m_{\bar q}][2xx'(1-x)(1-x')m_{Q_1} \nonumber\\
&&+(1-x)(1-x')(x+x'-2xx')m_{\bar q}+x'(1-x)(x-x')m_{Q_2}] \nonumber\\
&&+\frac{k^{2}_{\bot}}{W_{V}}\left[2xx'(1-x)(1-x')m_{Q_1}
        +2xx'(1-x)(1-x')m_{Q_2}\right]\bigg\} \, , \nonumber\\
A_2&=&\frac{2(x-x')\Theta_{P} k^{2}_{\bot}}{xx'^{2}(1-x')^{2}(1-x)^{2}}
\bigg\{k^{2}_{\bot}(x'+x-2x'x) \nonumber\\
&&+[x(1-x')m_{Q_1}-x'(1-x)m_{Q_2}][xm_{Q_1}-x'm_{Q_2}+(x'-x)m_{\bar q}]
 \nonumber\\
&&+(x+x'-2xx')[x'm_{Q_2}+(1-x')m_{\bar q}][xm_{Q_1}+(1-x)m_{\bar q}]
 \bigg\} \, ,
\ee
and
\be
\Theta_{P} &=& \frac{1}{\phi_{P}(x,k^2_{\bot})}
\frac{d\phi_{P}(x,k^2_{\bot})}{dk^2_{\bot}}\, .
\ee
At $p^2=0$ in which we have $x=x'$, it can be shown from Eq.~(\ref{f1r})
that
\be
F_{1}(0)&=&\int dx d^{2}k_{\bot}N_{V}\phi_{V}(x,k_{\bot})
N_{P}\phi^{*}_{P}(x,k_{\bot})\frac{4}{x^{2}(1-x)^{2}}
\bigg\{(1-x)k^{2}_{\bot} \nonumber\\
&&+[x m_{Q_2}+(1-x)m_{\bar q}][x m_{Q_1}+(1-x)m_{\bar q}]
+\frac{k^{2}_{\bot}}{W_{V}}x(m_{Q_1}+m_{Q_2})\bigg\}\, .
\label{48}
\ee
We note that Eq.~(\ref{48}) is the same as that in Ref. \cite{Don95}.

The calculations of the form factors $F_{2}$ and $F_{3}$ are more complex
than that of $F_1$ because they cannot be determined separately.
 From Eq.~(\ref{5}), we have that
\be
&&\langle V(q,\epsilon)|\bar{q}i\sigma^{+\nu}p_\nu \gamma_5 Q |P(p+q)\rangle \nonumber\\
&&~~~~~~=
\bigg\{-(1+2r) F_{2}
+\left[1-\frac{p^2}{M_{P}^{2}-M_{V}^{2}}(1+2r)
\right]F_{3}\bigg\}\frac{1}{r}(\epsilon_{\bot} \cdot q_{\bot})\,,  \label{f231}
\ee
and
\be
&& \langle V(q,\epsilon)|\bar{q}i\sigma^{+\nu}p_\nu \gamma_5
Q|P(p+q)\rangle \nonumber \\
&&~~~~~~=
\biggl[(m_{P}^{2}-m_{V}^{2})\frac{r}{M_{V}}
 -\frac{1+2r}{2M_{V}}\left(\frac{rM_{P}^{2}}{1+r}-M_{V}^{2}-\frac{M_{V}^{2}}{r}
\right)\biggr]F_{2} \nonumber\\
&&~~~~~~~~+ 
\frac{1}{2M_{V}}\left(\frac{rM_{P}^{2}}{1+r}-M_{V}^{2}-\frac{M_{V}^{2}}{r}
\right)\times \left[1-\frac{p^2}{M_{P}^{2}-M_{V}^{2}}(1+2r)
\right]F_{3}\,,  \label{f230}
\ee
for the transverse and longitudinal vector mesons polarizations of
 $\varepsilon^{\mu}(1)$ and $\varepsilon^{\mu}(0)$, respectively, where
\be
r=\frac{q^+}{p^+}=\frac
{-(p^2+M_{V}^{2}-M_{P}^{2})+\sqrt{(p^2+M_{V}^{2}-M_{P}^{2})^{2}
-4p^2 M_{V}^{2}}}{2p^2}\, .
\ee
Using Eqs.~(\ref{loop}), (\ref{p1}) and (\ref{p0}), we get
\be
&&\bigg\{-(1+2r) F_{2}
+\left[1-\frac{p^2}{M_{P}^{2}-M_{V}^{2}}(1+2r)
\right]F_{3}\bigg\}\frac{1}{r} \nonumber\\
&&~~~~~=\int dx' d^{2}k_{\bot}N_{V}\phi^{*}_{V}(x',k_{\bot})
N_{P}\phi_{P}(x,k_{\bot})(B_1+B_2)\equiv h(p^2)
\label{ans1}
\ee
and
\be
&&\biggl[(m_{P}^{2}-m_{V}^{2})\frac{r}{M_{V}}
-\frac{1+2r}{2M_{V}}\left(\frac{rM_{B}^{2}}{1+r}-M_{V}^{2}-\frac{M_{V}^{2}}{r}
\right)\biggr]F_{2} \nonumber\\
&&+\frac{1}{2M_{V}}\left(\frac{rM_{B}^{2}}{1+r}-M_{V}^{2}-\frac{M_{V}^{2}}{r}
\right)\left[1-\frac{p^2}{M_{P}^{2}-M_{V}^{2}}(1+2r)
\right]F_{3} \nonumber\\
&&~~~~~=\int dx' d^{2}k_{\bot}N_{V}\phi^{*}_{V}(x',k_{\bot})
N_{P}\phi_{P}(x,k_{\bot})C \equiv g(p^2)
\label{ans2}
\ee
where
\be
B_1&=&\frac{2}{xx'^{2}(1-x')^{2}(1-x)^{2}}\bigg\{
(2x'-2x'^{2}-x+xx')(x'+x-2xx')k^{2}_{\bot}\nonumber\\
&&-\frac{1}{W_{V}}\biggl[2xx'(1-x)(1-x')m_{Q_1}
        +2(1-x')(x'-x)(x'+x-2xx')m_{\bar q}\nonumber\\
&&+2x'(1-x')(x-x')^{2}m_{Q_2}\biggr]k^{2}_{\bot}
-\biggl([x'm_{Q_2}+(1-x')m_{\bar q}][2xx'(1-x)(1-x')m_{Q_1} \nonumber\\
&&+(1-x)(1-x')(x+x'-2xx')m_{\bar q}-x'(1-x)(x-x')m_{Q_2}]\biggr)\bigg\}\, ,
 \nonumber\\
B_2&=&\frac{2(x-x')\Theta_P k^{2}_{\bot}}{xx'^{2}(1-x')^{2}(1-x)^{2}}\bigg\{
        (2x'-1)(x'+x-2xx')k^{2}_{\bot}\nonumber\\
&&-\frac{2}{W_{V}}\biggl[x(1-x)(2x'-1)m_{Q_1}
+(x'-x)(x'+x-2xx')m_{\bar q}\nonumber\\
&&+x'(1-x')(2x-1)m_{Q_2}\biggr]k^{2}_{\bot}
+\biggl[[x(1-x')m_{Q_1}-x'(1-x)m_{Q_2}] \nonumber\\
&&\times [(2x'-1)m_{Q_1}+x'm_{Q_2}+(x+x'-2xx')m_{\bar q}] \nonumber\\
&&-(x+x'-2xx')[x'm_{Q_2}+(1-x')m_{\bar q}][xm_{Q_1}+(1-x)m_{\bar q}]
\nonumber\\
&&-\frac{2}{W_{V}}\biggl[x^2 x'(1-x')m_q m_{b}^{2}
-x^{2}(1-x')^{2}m_{\bar q} m_{Q_1}^{2}\nonumber\\
&&-x(1-x)(1-x')^{2}m_{Q_1} m_{\bar q}^{2}
        +xx'^2 (1-x)m_{Q_1} m_{Q_2}^{2}\nonumber\\
&&+x'^{2}(1-x)^{2}m_{\bar q} m_{Q_2}^{2}
-x'(1-x)^{2}(1-x')m_{Q_2} m_{\bar q}^{2}\biggr]\bigg\}\, , \nonumber\\
C&=&\frac{2}{xx'^{2}(1-x')^{2}(1-x)^{2}M_{0V}}
    \bigg\{\biggl[x'(1-x')M_{0V}^{2}+k^{2}_{\bot}+m_{Q_2} m_{\bar q}\biggr]
        k^{2}_{\bot}(x'+x-2xx')\nonumber\\
&&
+\biggl[x'(1-x')M_{0V}^{2}+k^{2}_{\bot}+m_{Q_2} m_{\bar q}\biggr]
\nonumber\\ &&\times
        (x m_{Q_1}+(1-x)m_{\bar q})(x'(1-x)m_{Q_2}+x(1-x')m_{Q_1})\nonumber\\
&&+\frac{1}{2W_{V}}\biggl[\frac{m_{Q_2}^{2}+k^{2}_{\bot}}{1-x'}
        -\frac{m_{\bar q}^{2}+k^{2}_{\bot}}{x'}-(1-2x')M_{0V}^{2}\biggr]
\nonumber\\
&&\times (xm_{Q_1}+(1-x)m_{\bar q})(x'm_{Q_2}-(1-x')m_{\bar q})
(x'(1-x)m_{Q_2}+x(1-x')m_{Q_1})\nonumber\\
&&-(m_{Q_2}-m_{\bar q})
\nonumber\\ &&\times\biggl(x(1-x)(2x'-1)m_{Q_1}
        +(1-x)(x'+x-2xx')m_{\bar q}
-x'(1-x)m_{Q_2}\biggr)k^{2}_{\bot}\nonumber\\
&&+\frac{1}{2W_{V}}\biggl[\frac{m_{Q_2}^{2}+k^{2}_{\bot}}{1-x'}
        -\frac{m_{\bar q}^{2}+k^{2}_{\bot}}{x'}-(1-2x')M_{0V}^{2}\biggr]
\nonumber\\
&&\times \biggl(x'(1-x')(2x-1)m_{Q_2}+(x'-x)(x'+x-2xx')m_{\bar q}
\nonumber\\
&&+x(1-x)(2x'-1)m_{Q_1}\biggr)k^{2}_{\bot}\bigg\}\,
.
\ee
The physical kinematic range of $p^2 :0 \to (M_{P}-M_{V})^2$ corresponds to
\be
\frac{x}{x'}\, : 1 \to \frac{M_{V}}{M_{P}}\,.
\ee
 From Eqs.~(\ref{ans1}) and (\ref{ans2}), it is not difficult to find
$F_2$ and $F_3$ as follows:
\be
(M_{P}^{2}-M_{V}^{2})F_2&=&\frac{M_V}{r}h(p^{2})-\frac{g(p^2)}{2}
\biggl(\frac{r}{1+r}M_{P}^{2}-\frac{1+r}{r}M_{V}^{2}\biggr)\, , \nonumber\\
\biggl[1-\frac{p^2}{M_{P}^{2}-M_{V}^{2}}\biggr]F_3&=&
\frac{M_V}{M_{P}^{2}-M_{V}^{2}}\frac{1+2r}{r}h(p^{2})\nonumber\\
&&+g(p^2)\biggl[r-\frac{1}{M_{P}^{2}-M_{V}^{2}}\frac{1+2r}{2}
\biggl(\frac{r}{1+r}M_{P}^{2}-\frac{1+r}{r}M_{V}^{2}\biggr)\biggr]\, .
\label{f23}
\ee
At the maximum recoil of $p^2=0$,
$F_2(0)$ and $F_3(0)$ are simply given by
\be
F_2(0)&=&-\int dx d^{2}k_{\bot}N_{V}\phi^{*}_{V}(x,k_{\bot})
N_{P}\phi_{P}(x,k_{\bot})\nonumber\\
&&\frac{4}{x^{2}(1-x)^{2}}\bigg\{
xk^{2}_{\bot}-\frac{k^{2}_{\bot}}{W_{V}}x m_{Q_1}
-[x m_{Q_2}+(1-x)m_{\bar q}][x m_{Q_1} +(1-x)m_{\bar q}]\bigg\}\,, \nonumber\\
F_3(0)&=&\int dx d^{2}k_{\bot}N_{V}\phi^{*}_{V}(x,k_{\bot})
N_{P}\phi_{P}(x,k_{\bot})\nonumber\\
&&\frac{M_V}{M_{P}^{2}-M_{V}^{2}} \frac{2}{x^{2}(1-x)^{3}M_{0V}}
\bigg\{\biggl[x(1-x)M_{0V}^{2}+k^{2}_{\bot}+m_{Q_2} m_{\bar q}\biggr]
\nonumber\\
&&\times [(x m_{Q_1}+(1-x)m_{\bar q})(m_{Q_2}+m_{Q_1})+2k^{2}_{\bot}]
\nonumber\\
&&+\frac{m_{Q_2}+m_{Q_1}}{2W_{V}}\biggl[\frac{m_{Q_2}^{2}+k^{2}_{\bot}}{1-x}
-\frac{m_{\bar q}^{2}+k^{2}_{\bot}}{x}-(1-2x)M_{0V}^{2}\biggr]\nonumber\\
&&\times [(xm_{Q_1}+(1-x)m_{\bar q})(xm_{Q_2}-(1-x)m_{\bar q})
+(2x-1)k^{2}_{\bot}] \nonumber\\
&&-(m_{Q_2}-m_{\bar q})
\biggl((2x-1)m_{Q_1}+2(1-x)m_{\bar q}-m_{Q_2}\biggr)k^{2}_{\bot}
\bigg\}+\frac{F_2(0)}{2}\,.
\ee
Thus, we have evaluated all the tensor form factors of
$F_T$, $F_1$, $F_2$ and $F_3$ in the whole physical region of
the momentum transfer with the light front quark model.

\se{Numerical Results}

\ \ \

We now calculate numerically 
the tensor form factors
in the LFQM.
The parameters in the light front wave functions are fixed by
other hadronic properties.
We use the decay constants in Eqs.~(\ref{fp}) and (\ref{fv}) to determine the
parameter $\omega$.
The measured decay constants of light pseudoscalar and vector mesons are
\be
f_\pi=\,132\,{\rm MeV},~~f_\rho=\,216\,{\rm MeV},~~f_K=\,160\,{\rm MeV},
~~f_{K^*}=\,210\,{\rm MeV}.
\label{fpl}
\ee
Since the decay constants of heavy mesons are unknown experimentally,
we have to rely on various calculations in QCD-motivated models \cite{qcd1,qcd2,qcd3}.
We shall take
\be
f_B=\,185\,{\rm MeV}\,,~~~f_{B^*}=\,205\,{\rm MeV}\,.
\label{fph}
\ee
In Table 1, we list the parameters $\omega_{M}~(M=\pi,~\rho,~K,~K^{*},~B$
and $B^{*})$
 fitted by the decay constants in Eqs.~(59) and (60) with the
Gaussian-type wave functions.
Note that the quark masses given in
Table 1 are chosen to the commonly used values in relativistic quark
models.

\begin{table}
\caption{Parameters $\omega_{M}~(M=\pi,~\rho,~K,~K^{*},~B$ and
$B^{*})$.}
\begin{center}
\begin{tabular}{|c||c c c|c c c|c c c|} \hline
~wave function~ & $m_{u,d}$ & $\omega_\pi$ & $\omega_\rho$ & $m_s$ &
$\omega_K$
& $\omega_{K^*}$ & $m_b$ & $\omega_B$
& $\omega_{B^*}$ \\  \hline
Gaussian & ~0.25 & 0.33 & 0.31~ & ~0.40 & 0.38 & 0.31~
& ~4.8 & 0.55 & 0.55~\\ \hline
\end{tabular}
\end{center}
\end{table}

We now show the form factors $F_T$, $F_1$, $F_2$ and $F_3$ in the entire
physical region $0\leq p^2\leq (M_i - M_f)^2$.
The $p^2$ dependences of $F_T$
for $B \to K$ and $B \to \pi$ are shown in Figures~1 and 2, respectively.
  From Figure~1, one can see
that $F_T$ will decrease as $p^2$ increases near the zero recoil
point. This is reasonable because the matrix elements depend on the
overlap integral of the initial and final meson wave functions.
For the heavy  to light transition, the internal momentum distribution of
the heavy
meson $\phi(x,k_\bot)$ has a narrow peak near $x=0$, whereas the peak of
$\phi(x,k_\bot)$ for the light meson has a larger width than
that for the heavy one.
Therefore, the maximum overlapping of these two wave functions
occurs away from the zero recoil point. At the maximum recoil of $p^2=0$,
we have the values
\be
F_T^{B\pi}(0)=0.27\,,~~~~F_T^{BK}(0)=0.36\, . \label{fpvalue}
\ee
We note that he values in Eq.~(61) are close to those in the light cone
QCD sum rule models
\cite{ffpp1,ffpp2,ffpp3}.

The form factors $F_{1,2,3}$ as a function of $p^2$ for
$B\to K^{*}$ and $B\to \rho$ transitions are depicted in Figures~3 and 4,
respectively.
At $p^2=0$, we have
\be
&& B\to K^*:~~~~F_1^{BK^*}(0)=0.63\,,~~~F_2^{BK^*}(0)=0.32\,,~~~F_3^{BK^*}(0)=
0.21\, \nonumber\\
&& B\to \rho:~~~~F_1^{B\rho}(0)=0.54\,,~~~F_2^{B\rho}(0)=0.27\,,
~~~F_3^{B\rho}(0)=0.19\, .
\ee

To compare our results with those in the literature, we may fit approximately
the $p^2$ behaviors of the form factors in Figures~1-4 as
pole-like forms:
\be
F_{i}(p^2)=\frac{F_{i}(0)}{1-p^2 / \Lambda_{1}^{2}+p^4 / \Lambda_{2}^{4}}\,~~~
(i=T,~1,~2,~3)
\ee
where the values of $\Lambda_{1}$ and $\Lambda_{2}$ are listed in the
Table 2 with $F_T$ for $B \to \pi,~K$ and $F_{1,2,3}$ for $B \to
\rho,~K^{*}$.

\begin{table}
\caption{The values of $\Lambda_{1}$ and $\Lambda_{2}$ for the
corresponding $F_{i}(0)$ in $B \to M$~
$(M=\pi,~K,~\rho,~K^{*})$.}
\begin{center}
\begin{tabular}{| c || c || c || c | c | c || c | c | c |} \hline
$M$ & $\pi$ & $K$ & \multicolumn{3}{c||}{$\rho$} & 
\multicolumn{3}{c|}{$K^{*}$}  \\ \hline
$F_{i}(p^{2})$ & $F_T$ & $F_T$ & $F_1$ & $F_2$ & $F_3$
& $F_1$ & $F_2$ & $F_3$  \\ \hline
$p^{2}=0$ & 0.27 & 0.36 & 0.54 & 0.27 & 0.18 & 0.63 & 0.32 & 0.21  \\ \hline
$\Lambda_{1}({\rm GeV})$ & 4.32 & 4.36 & 4.03 & 6.04 & 4.45 & 3.96 
& 6.17 & 4.78  \\ \hline
$\Lambda_{2}({\rm GeV})$ & 5.38 & 5.29 & 5.43 & 12.45 &
5.84 & 5.18 & 11.02 & 6.17  \\ \hline
\end{tabular}
\end{center}
\end{table}

As shown in Table 3, we see that our results of $F_{i}(0)$ agree well
with those in the literature
\cite{Don95,ffpv,lattice}. We remark that the relations among the form factors
for the $P \to V$ transition in Ref. \cite{rel} can be tested in our
calculations.
We find that the relation of
Eqs.~(5)-(7) in Ref. \cite{rel} are well satisfied at $p^2 \to 0$ and hold
with 20\% accuracy at $p^2 \leq 10 \ \ GeV^2$. However, they are violated for
$p^2$ larger than $10\ \ GeV^2$. Finally, we note that the relation
$F_1(0) = 2~F_2(0)$ \cite{ffpv}
is satisfied and $F_1(p^2)$ increases as $p^2$ faster than $2~F_2(p^2)$.

\begin{table}
\caption{Comparison of different works for the tensor form
factors at $p^2=0$}
\begin{center}
\begin{tabular}{|c||c |c |c |c |} \hline
$B\to \pi(\rho)$ & This work &  \cite{Don95} & \cite{ffpv} & \cite{lattice} \\
\hline
$F_T(0)$ & 0.27 & - & - & - \\ \hline
$F_1(0)$ & 0.54 & 0.56 & $0.58\pm 0.08$ & - \\ \hline
$F_2(0)$ & 0.27 & - & $0.29\pm 0.04$ & - \\ \hline
$F_3(0)$ & 0.19 & - & $0.20\pm 0.03$ & - \\ \hline   \hline
$B\to K(K^*)$ & This work &  \cite{Don95} & \cite{ffpv} & \cite{lattice} \\
\hline
$F_T(0)$ & 0.36 & - & - & - \\ \hline
$F_1(0)$ & 0.63 & 0.74 & $0.76\pm 0.12$ & $0.64^{+0.08}_{-0.04}$ \\ \hline
$F_2(0)$ & 0.32 & - & $0.38 \pm 0.06$ & $0.32^{+0.04}_{-0.02}$  \\ \hline
$F_3(0)$ & 0.21 & - & $0.26 \pm 0.04$ & - \\ \hline
\end{tabular}
\end{center}
\end{table}

\se{Conclusions}

\ \ \

The tensor form factors of $F_T$ and $F_{1,2,3}$ for $P \to P$ and $P \to V$
transitions have been studied in the LFQM. These form factors are
 evaluated in the entire physical momentum transfer region of $0\leq
p^2\leq (M_i - M_f)^2$.
 We have used the values of the decay constants and the constitute quark
masses
to fix the parameters $\omega_{M}$ appearing in the wave functions. Thus, there
are no more degree of freedom to adjust the light front wave functions.
We have fitted our numerical results of the form factors
in $B\to \pi,K,\rho,K^*$
 into dipole forms and
we have shown that our results agree well with those in the literature. \\

\noindent
{\bf Acknowledgments}

This work was supported in part by the National Science Council of the
Republic of China under the Grant Nos. NSC89-2112-M-007-054 and
NSC-89-2811-M-006-013 and the National Center for Theoretical Science.

\newpage

\begin{figcap}

\item
The values of $F_{T}$ as a function of the
momentum transfer $p^2$ for $B\to K$.
\item
Same as Figure 1 but for $B\to \pi$.
\item
The values of $F_{1,2,3}$ as a function of the
momentum transfer $p^2$ for $B\to K^*$.
\item
Same as Figure 3 but for $B\to \rho$.

\end{figcap}

\newpage
\begin{figure}[h]
\includegraphics{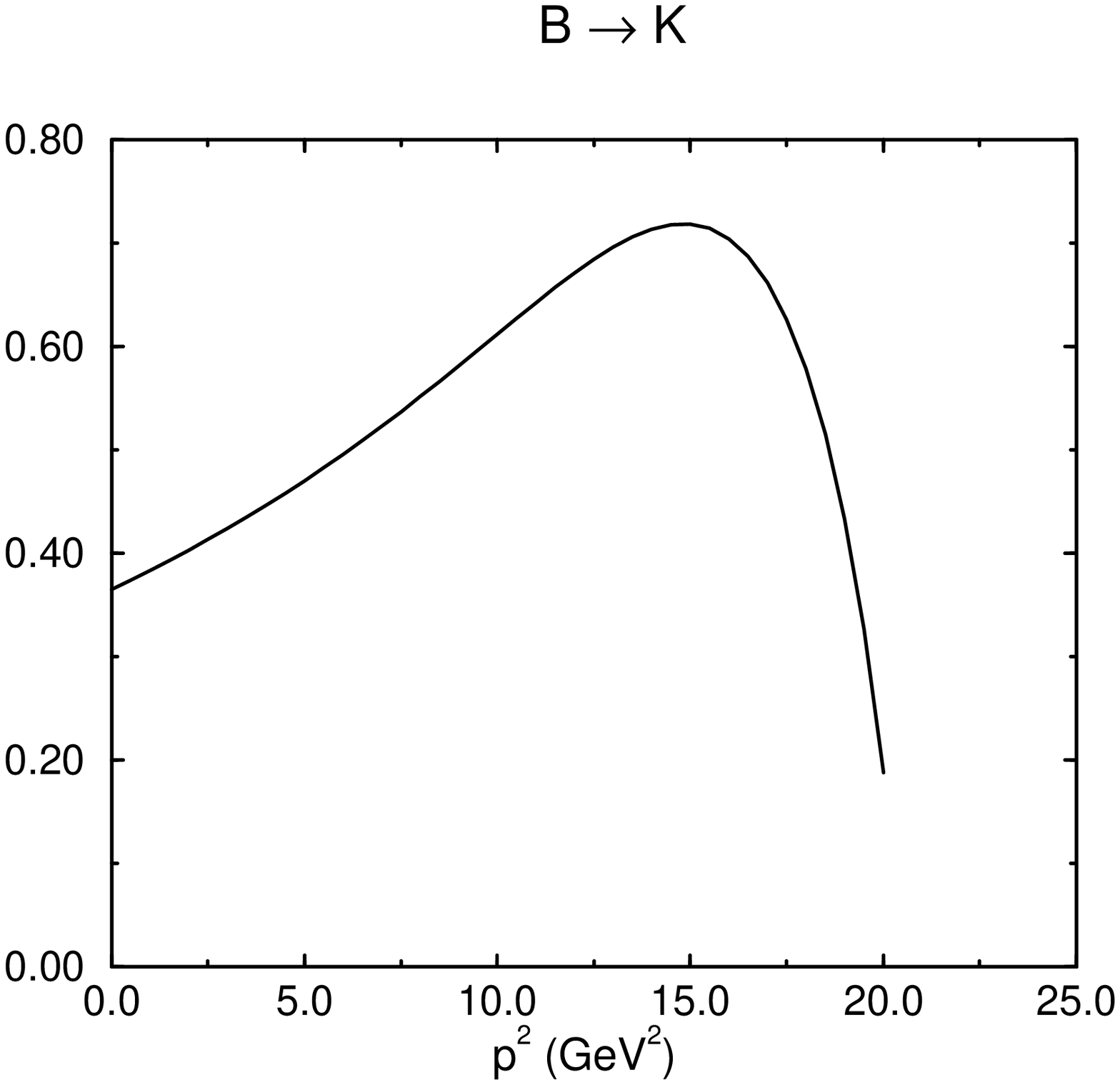}
\vskip 13cm
\caption{
The values of the form factors $F_{T}$ as functions of the
momentum transfer $p^2$ for $B\to K$.
}
\end{figure}

\newpage
\begin{figure}[h]
\includegraphics{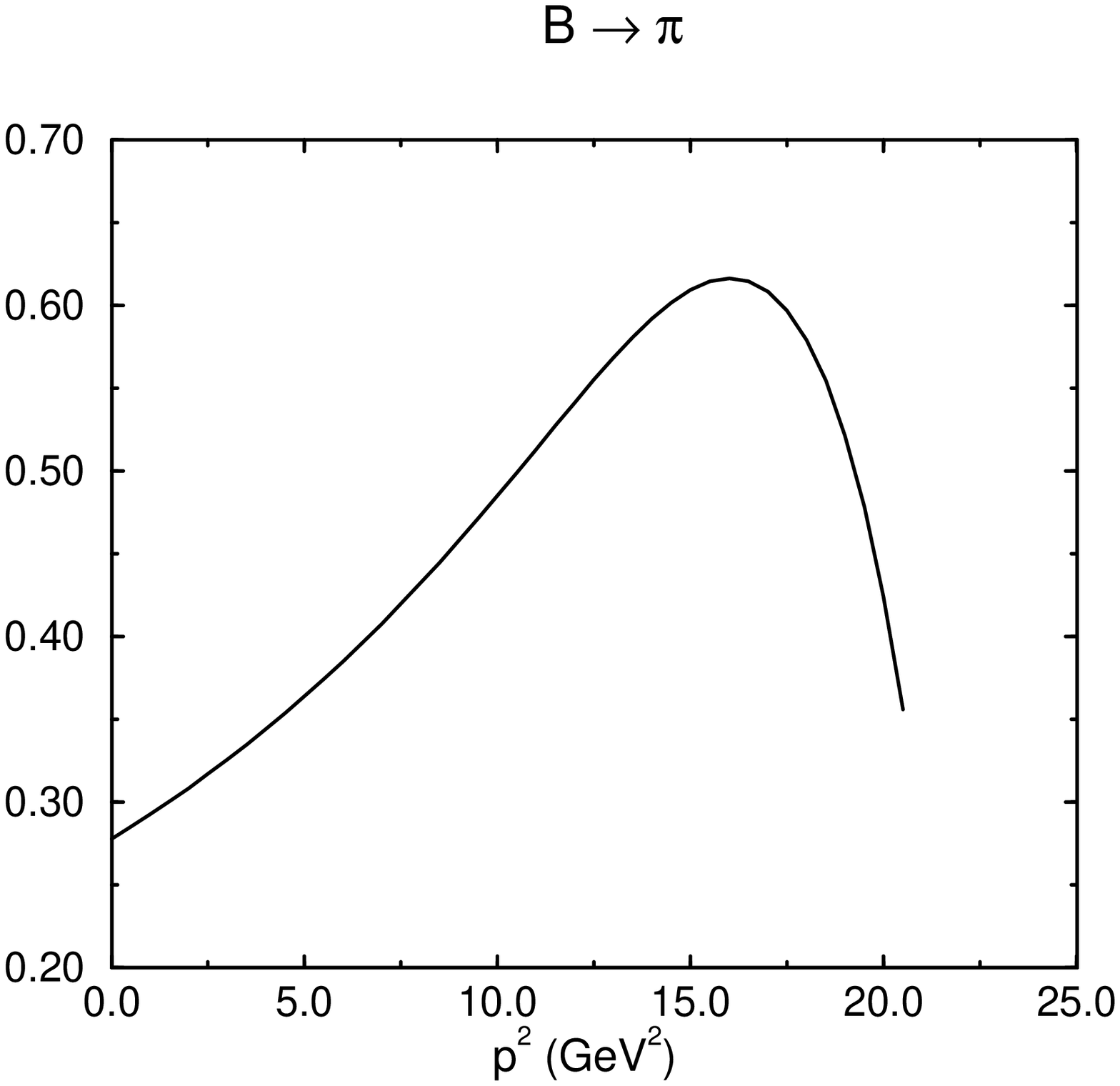}
\vskip 13cm
\caption{
The values of the form factors $F_{T}$ as functions of the
momentum transfer $p^2$ for $B\to \pi$.
}
\end{figure}

\newpage
\begin{figure}[h]
\includegraphics{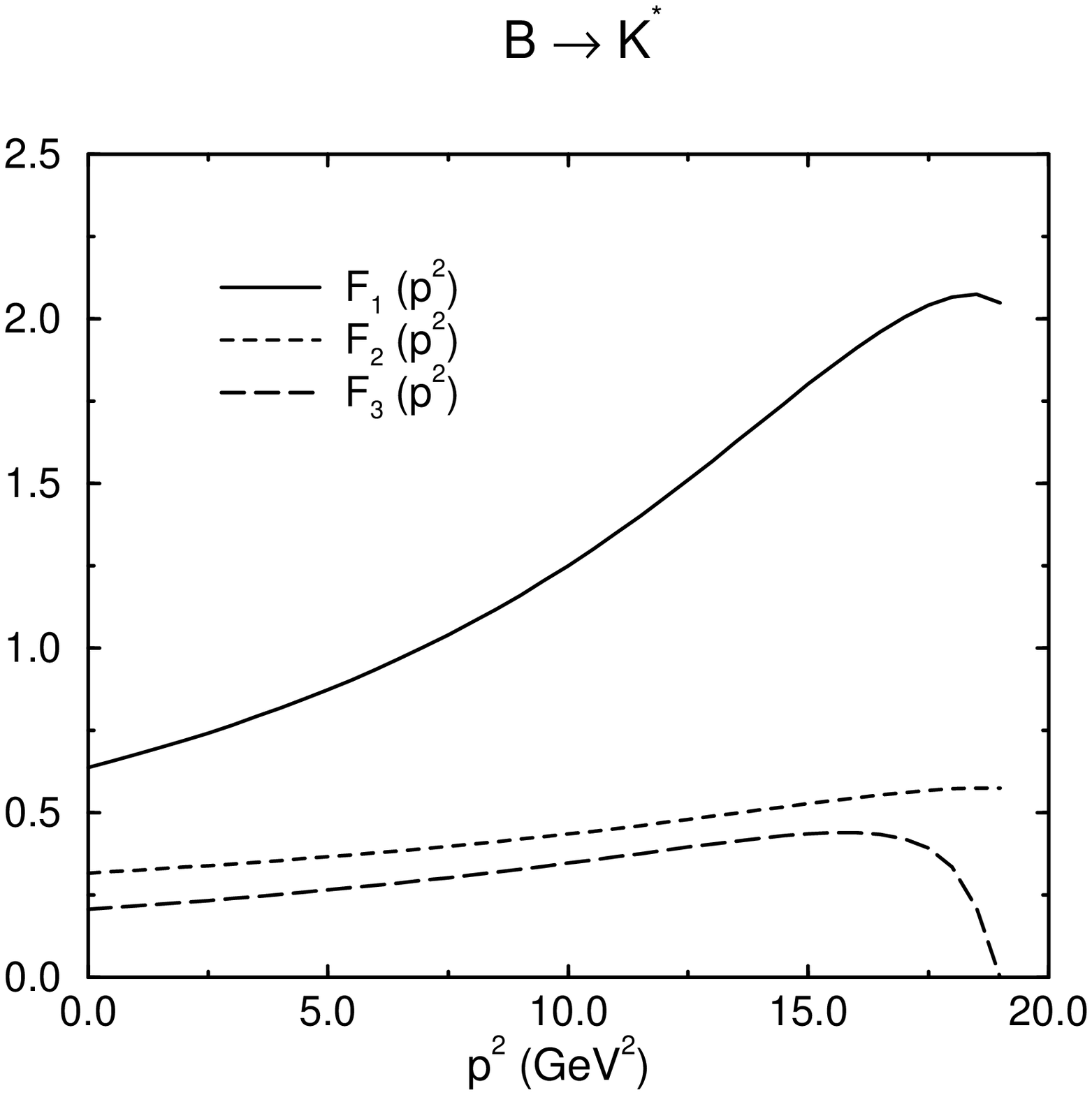}
\vskip 13cm
\caption{The values of the form factors $F_{1,2,3}$ as functions of the
momentum transfer $p^2$ for $B\to K^*$.}
\end{figure}

\newpage
\begin{figure}[h]
\includegraphics{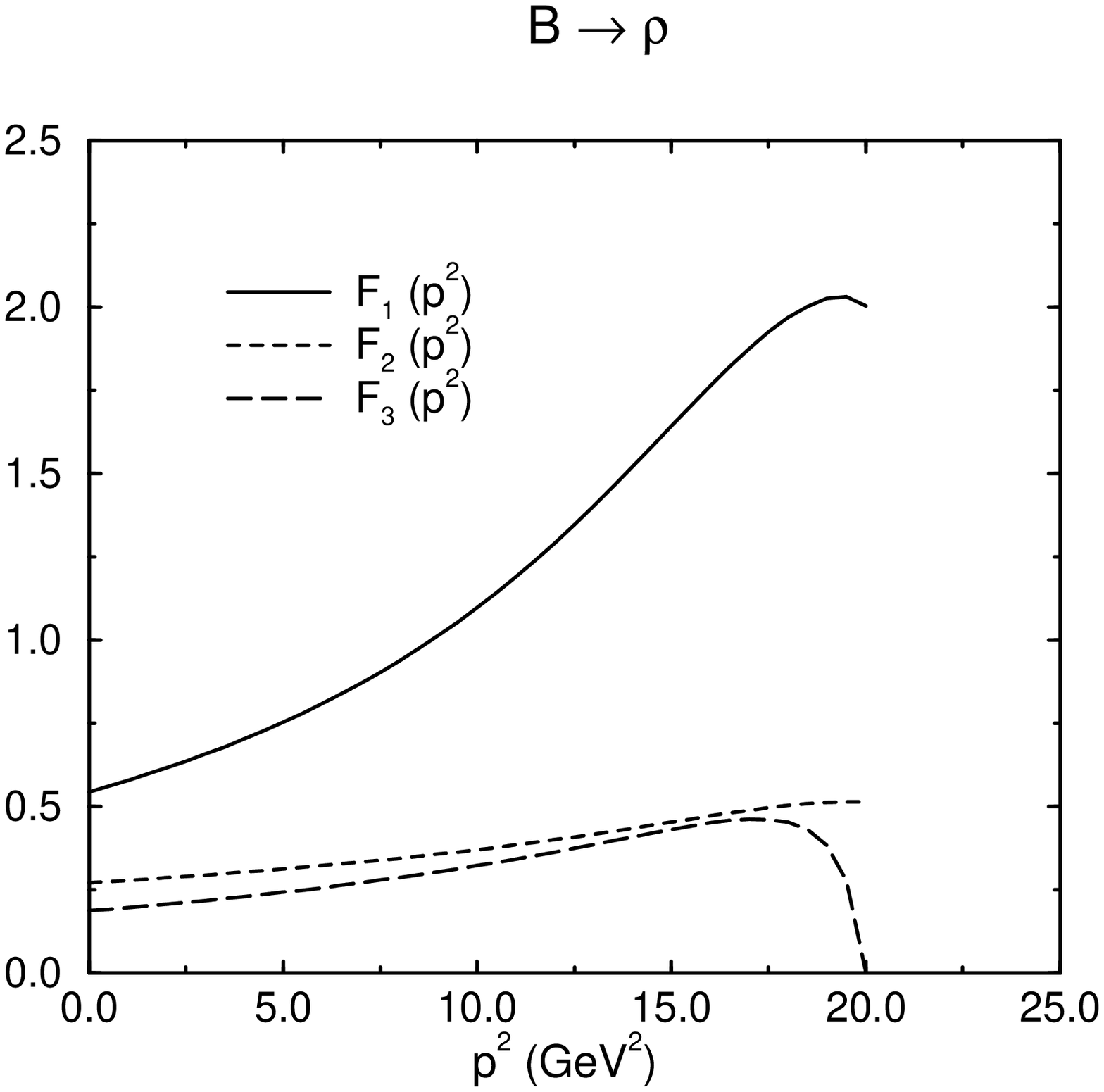}
\vskip 13cm
\caption{The values of the form factors $F_{1,2,3}$ as functions of the
momentum transfer $p^2$ for $B\to \rho$.}
\end{figure}

\end{document}